\documentclass[]{spie}  %>>> use for US letter paper
\usepackage{amsmath,amsfonts,amssymb}
\usepackage{graphicx}
\usepackage[colorlinks=true, allcolors=blue]{hyperref}

\title{Sketch-Based 3D Shape Modeling from Sparse Point Clouds}

\author[a]{Xusheng Du}
\author[a]{Yi He}
\author[b]{Xi Yang}
\author[c]{Chia-Ming Chang}
\author[a]{Haoran Xie}
\affil[a]{Japan Advanced Institute of Science and Technology, Ishikawa, Japan}
\affil[b]{Jilin University, Changchun, China}
\affil[c]{The University of Tokyo, Tokyo, Japan}

\authorinfo{Further author information: (Send correspondence to H. Xie)
\\X. Du: E-mail: duxusheng@jaist.ac.jp
\\Y. He: E-mail: s2010035@jaist.ac.jp
\\X. Yang: E-mail: earthyangxi@gmail.com
\\C. M. Chang: E-mail: chiaming@ui.is.s.u-tokyo.ac.jp
\\H. Xie: E-mail: xie@jaist.ac.jp, Telephone: 81-0716-51-1753}

\begin{document} 
\maketitle

\begin{abstract}
3D modeling based on point clouds is an efficient way to reconstruct and create detailed 3D content. However, the geometric procedure may lose accuracy due to high redundancy and the absence of an explicit structure. In this work, we propose a human-in-the-loop sketch-based point cloud reconstruction framework to leverage users’ cognitive abilities in geometry extraction. We present an interactive drawing interface for 3D model creation from point cloud data with the help of user sketches. We adopt an optimization method in which the user can continuously edit the contours extracted from the obtained 3D model and retrieve the model iteratively. Finally, we verify the proposed user interface for modeling from sparse point clouds.
\end{abstract}


\section{INTRODUCTION}
\label{sec:intro}  % \label{} allows reference to this section

With the rapid development of 3D modeling technology, 3D models are currently used in various applications, such as in the medical industry, films, video games, and architectural designs. It is increasingly important to retrieve and reconstruct 3D models from sparse input. 3D data can be represented in different formats depending on the end use. Common formats include depth maps, point clouds, voxels, meshes, and volumetric grids. In particular, point clouds can preserve high-quality geometric information in 3D space without discretization. Therefore, it is a feasible representation for various real-world applications using point clouds, which can be easily obtained using 3D imaging sensors such as binocular cameras, depth cameras, and LiDAR laser scanning.

\begin{figure} [t]
\begin{center}
\begin{tabular}{c} 
\includegraphics[height=3cm]{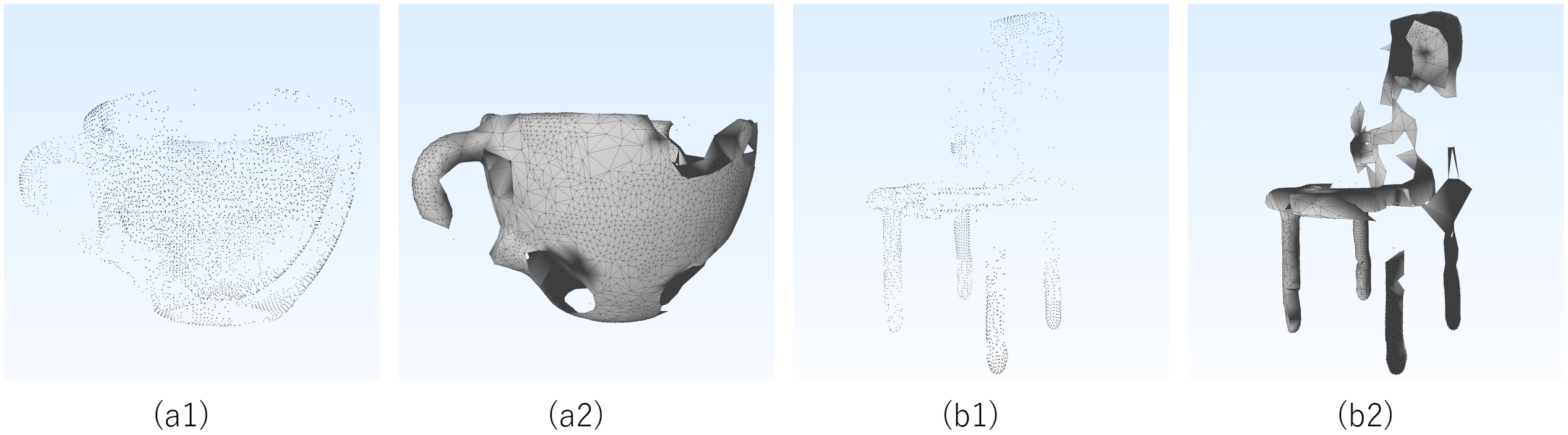}
\end{tabular}
\end{center}
\caption[example] 
{ \label{fig:example} 
The modeling results from sparse point clouds with the conventional surface reconstruction approach: (a1) the sparse point cloud data of a teacup and (a2) the reconstruction result; (b1) the sparse point cloud data of a chair and (b2) the reconstruction result. }
\end{figure}

However, the acquired point clouds may be noisy or incomplete with sparse density or partially missing data due to the limitations of the data acquisition process. Therefore, it is challenging to reconstruct and acquire high-quality 3D content efficiently from point clouds. Figure \ref{fig:example} shows the modeling results of MeshLab's automatic surface reconstruction approach. We observed that automatic processing is quite susceptible to the quality of point clouds. Previous work can reconstruct 3D models by recovering shapes from sketch inputs \cite{b1}, but the geometric shapes extracted from point clouds are usually simple and regular, and it is difficult to generate accurate 3D contents from sparse point clouds. Automatic extraction approaches may require pre-processes to simplify data or fill in missing data, which will complicate the process. Because users may have excellent cognitive abilities in spatial recognition, we represent the extraction of object shapes from point clouds as a human-in-the-loop tracing process.

In this work, we propose an interactive drawing interface to construct 3D models from point clouds, especially sparse point clouds. Our main contributions include the following: (i) an interactive 3D object drawing interface to handle sparse point cloud data and (ii) an iterative optimization method based on model contours with sketch retrieval and model contour editing on public point cloud datasets.

\section{Related Works}

Automatic shape modeling from point clouds has been explored extensively for various applications \cite{b17}.  PolyFit was proposed for reconstructing lightweight polygonal surfaces from point clouds \cite{b15}. Topological priors were recently incorporated into point scans for reconstructing point cloud surfaces \cite{b16}.
The automatic modeling approaches show robust and satisfying  reconstruction results for dense point clouds. However, these approaches may fail to provide a feasible solution for sparse point clouds, and we found that the human factor can greatly improve modeling results due to users' spatial cognitive abilities. 

Sketch-based user interface is an important research topic in computer graphics to improve automatic processes with human power, such as shape retrieval \cite{b8}, motion retrieval \cite{peng21}, image generation \cite{he2021,dualface2021} and task guidance \cite{xie21}. For image-based shape editing, 3-sweep introduced an interactive technique based on extracting simple 3D shapes from a single image \cite{b3}. Similarly, Gemketch can extend 3-sweep to extract 3D models of generalized rectangles and cylinders from single or multiple viewpoints \cite{b4}. However, these approaches usually require strong assumptions or expertise on natural images. In this work, we use point cloud data as drawing background and allow users to sketch interactively in an iterative routine until satisfactory modeling is achieved.

\section{System Overview}
\label{sec:sections}

% In this work, we propose a sketch-based modeling interface to help users construct the modeling shapes from sparse point clouds with user sketches. We first construct the sparse point clouds dataset, and utilize a data-driven sketch retrieval and model retrieval approaches to explore the desired 3D meshes from user input.

\begin{figure} [ht]
\begin{center}
\begin{tabular}{c} 
\includegraphics[height=6.5cm]{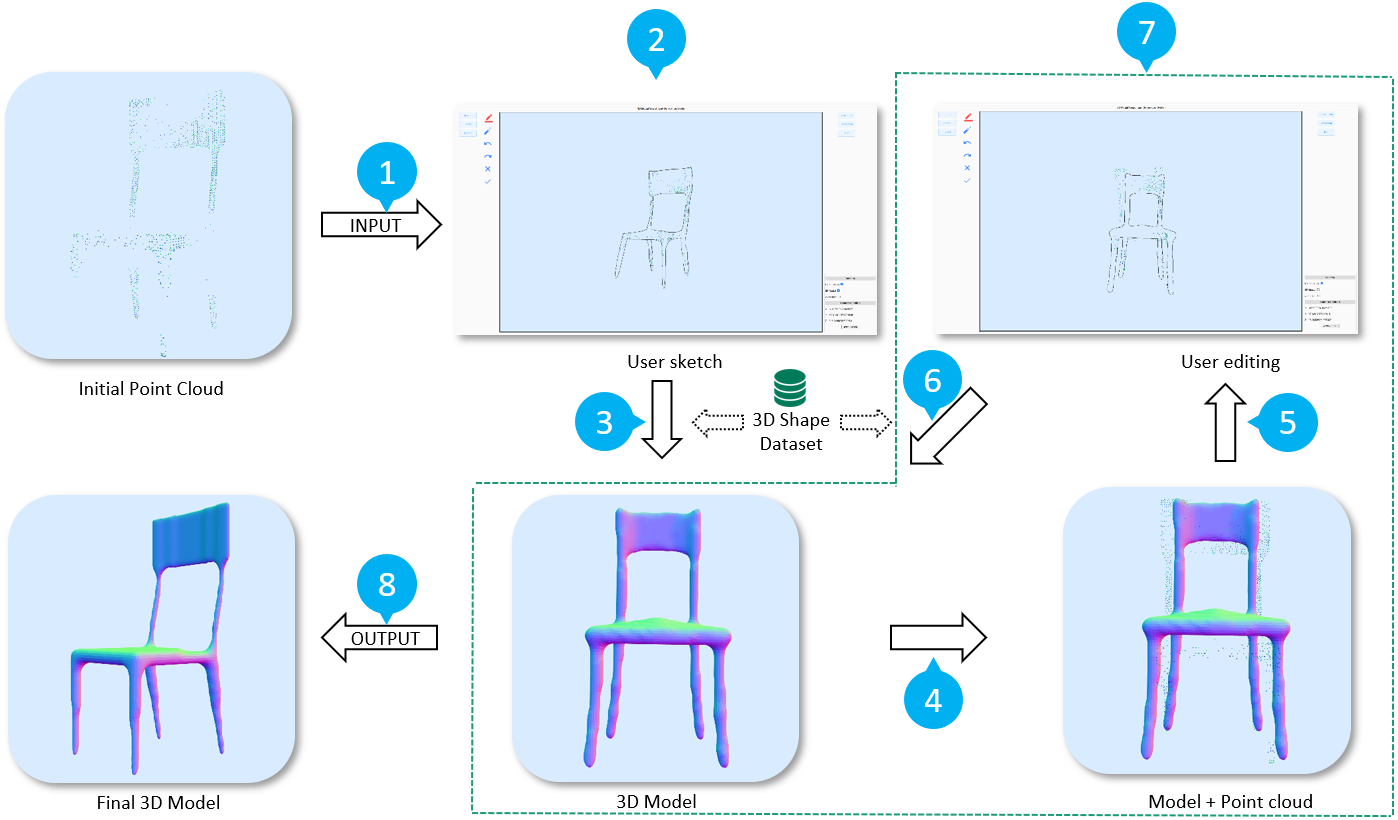}
\end{tabular}
\end{center}
\caption[example] 
{\label{fig:overview} 
System overview of the proposed model extraction with eight steps: (1) input of point cloud; (2) user sketch; (3) sketch retrieval of 3D models; (4) overlay with 3D model; (5) extraction of model contours for redrawing; (6) model re-retrieval; (7) repeat steps (4)—(6); (8) obtain the final model.}
\end{figure}

Figure \ref{fig:overview} shows the system overview of the proposed system. In the user interface, the user can adjust the angle to observe the spatial structure of the input sparse point cloud. After the observation, the user can determine any satisfactory angle and draw a sketch with reference to the sparse point cloud of the current angle. We overlay the sparse point cloud data with the retrieved 3D model with model alignment. If the user is not satisfied with the current model, the proposed interface allows the user to continue the retrieval based on the current model without  redrawing. After the user confirms that the sketch modification is complete, the user can perform a re-retrieval in an interactive manner to obtain a new 3D model. By extracting the contours of the retrieved model (step 5), users can  edit the contours and perform re-retrieval (step 6) in an interactive manner to continuously improve the overlap between the retrieved model and the initial point clouds.

\subsection{Datasets}
The dataset is constructed based on SHREC 2012 \cite{dataset12}. In our prototype design, our sparse point cloud database contained 100 point clouds, 20 for each of the 5 categories (teacup, chair, table, vase, and animal). In addition, we obtained 102 viewpoint matrixes from the 102 uniformly distributed view directions, and used these matrixes to capture contour images for each model. After these processes, we constructed a contour dataset of model data with 10,200 contour images. With the sketch input, the user can retrieve the most similar models from the dataset. We adopted OpenSSE \cite{b20} to retrieve the 3D model by comparing the similarity between the current sketch and the model contours in the contour dataset.

\subsection{Contour Extraction}
The extraction of model contours is mainly implemented based on OpenCV. First, we store the 3D model data of a specific angle displayed in the current canvas window as an image. This image is then converted into a grayscale image. Then we perform a median filtering operation on the grayscale image. The filtered image is then converted into a binary map using a threshold function. Finally, we extract the contour lines of the current 3D model from this binary map. Once the contour lines are extracted, we draw the contour lines to a new image file with the same image size as the current canvas window size. This new image file of the model contour is finally displayed back to the user interface as an input sketch. After the above process, we complete the extraction of the current model contours to the interface, which gives the user a sketch of the model contours that can also be modified.

\subsection{Model Alignment}

In this research, we adopt point cloud alignment to integrate the input and retrieved result into a unified coordinate system with an iterative closest point (ICP) approach \cite{b21}. First, we randomly select 2000 points in the sparse point cloud as control points to improve the computational efficiency and alignment accuracy of the ICP algorithm. We calculate the Euclidean distance of the corresponding points in the 3D model data, rotation matrix, and displacement vector using the control points and their corresponding points. We apply the computed rotation matrix and displacement vector to the 3D model data iteratively until the convergence condition is satisfied or the maximum number of iterations is achieved. Finally, the final aligned 3D model is displayed as the background of the proposed user interface.

\subsection{User Interface}
Figure 3(a) shows the proposed user interface. First, we provide the basic operations on point clouds and model files: importing point cloud data, exporting model files, and drawing or canceling function settings. The middle part is the canvas where the user can observe the point cloud and model and draw the sketch. We also provide ``brush," ``eraser," and ``undo" functions that are commonly used in drawings. We provide specific functions, including retrieving 3D models, extracting 3D model contours, and ICP alignment of point clouds and models. In addition, we can show the spatial coordinates of the current viewpoint and control the display and hiding of the axis system, point cloud, and 3D model on the main canvas. 

\begin{figure} [t]
\begin{center}
\begin{tabular}{c} 
\includegraphics[height=4.5cm]{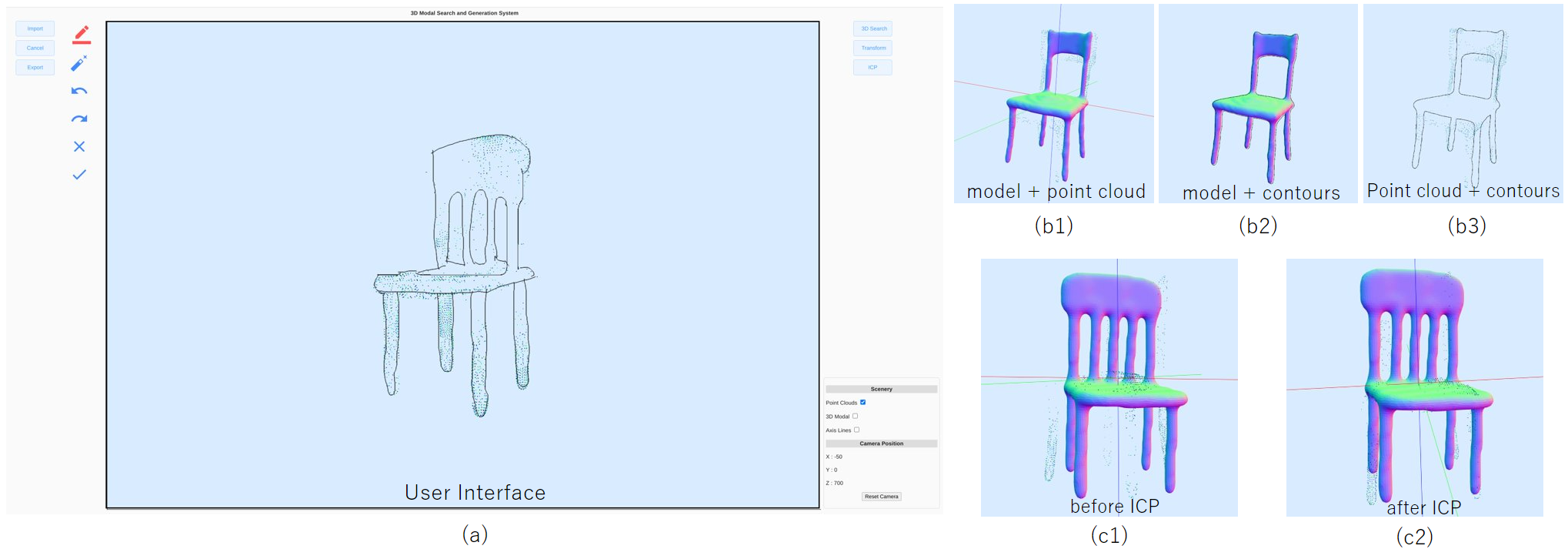}
\end{tabular}
\end{center}
\caption[example] 
{ \label{fig:interface} 
(a) The user is drawing sketches in the proposed user interface. (b1)—(b3) show the results before and after the model contour extraction, and (c1) and (c2) show the overlap of the point cloud and the model before and after ICP alignment, respectively.}
\end{figure}

\subsection{User Study}
We evaluated the proposed system quantitatively and qualitatively in user study. We invited six participants (three male and three female graduate students) to use the proposed interactive interface to extract 3D models from sparse point clouds. We quantitatively evaluated the results to compare the overlapping similarities of the point clouds and the retrieved models. In the user study, we recorded the number of times the user retrieved the target, the number of sketch modifications, and the ICP error value in the model alignment.

We asked all participants to conduct the designated experiment twice with the following three tasks: (1) select sparse point cloud data to input into MeshLab for surface reconstruction; (2) use our proposed system without inputting sparse point cloud data (the user draws a sketch randomly according to his own imagination, and then performs a database search based on that sketch to eventually verify if the target 3D model can be retrieved); and (3) use our proposed system with the input of sparse point clouds. For the sparse point clouds, the users draw the sketch according to their recognition of the spatial structure of that point cloud. Note that the users do not have prior knowledge of the target models. Users can perform sketch retrieval interactively until they believe that the feasible target 3D model is achieved.

\section{Results}

In this section, we present the modeling results using the proposed interface and the evaluation results from our user study to verify the effectiveness and user experience of our proposed interface.

\subsection{Modeling Results}

\begin{figure} [ht]
\begin{center}
\begin{tabular}{c} 
\includegraphics[height=3.5cm]{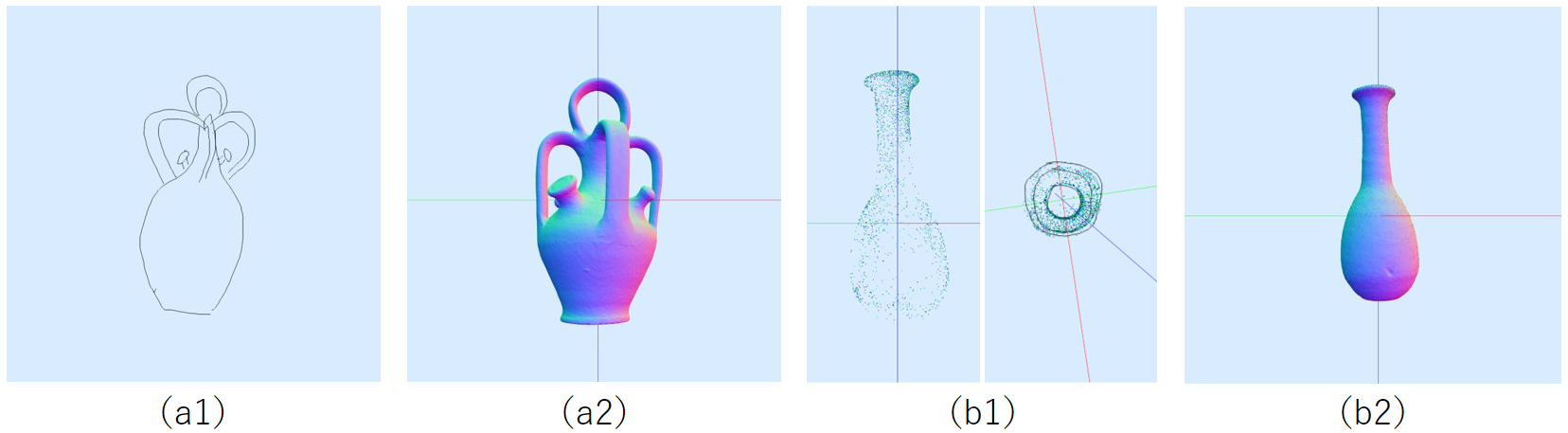}
\end{tabular}
\end{center}
\caption[example] 
{ \label{fig:result1} 
Sketch-based model retrieval result (a2) from the sketch drawn by the user freely  (a1), and 3D model result (b2) from the sparse point cloud of the vase only from the top view (b1.) }
\end{figure}

Our proposed system  can support not only 3D modeling from point clouds but also sketch-based model retrieval without point clouds. Figure \ref{fig:result1} shows that the user can input a sparse point cloud of a vase and retrieve the target model from only the top view. The proposed interface can involve the user's spatial cognitive ability so that users can sketch the spatial structure of the point cloud data successfully. 

\begin{figure} [ht]
\begin{center}
\begin{tabular}{c} 
\includegraphics[height=6.5cm]{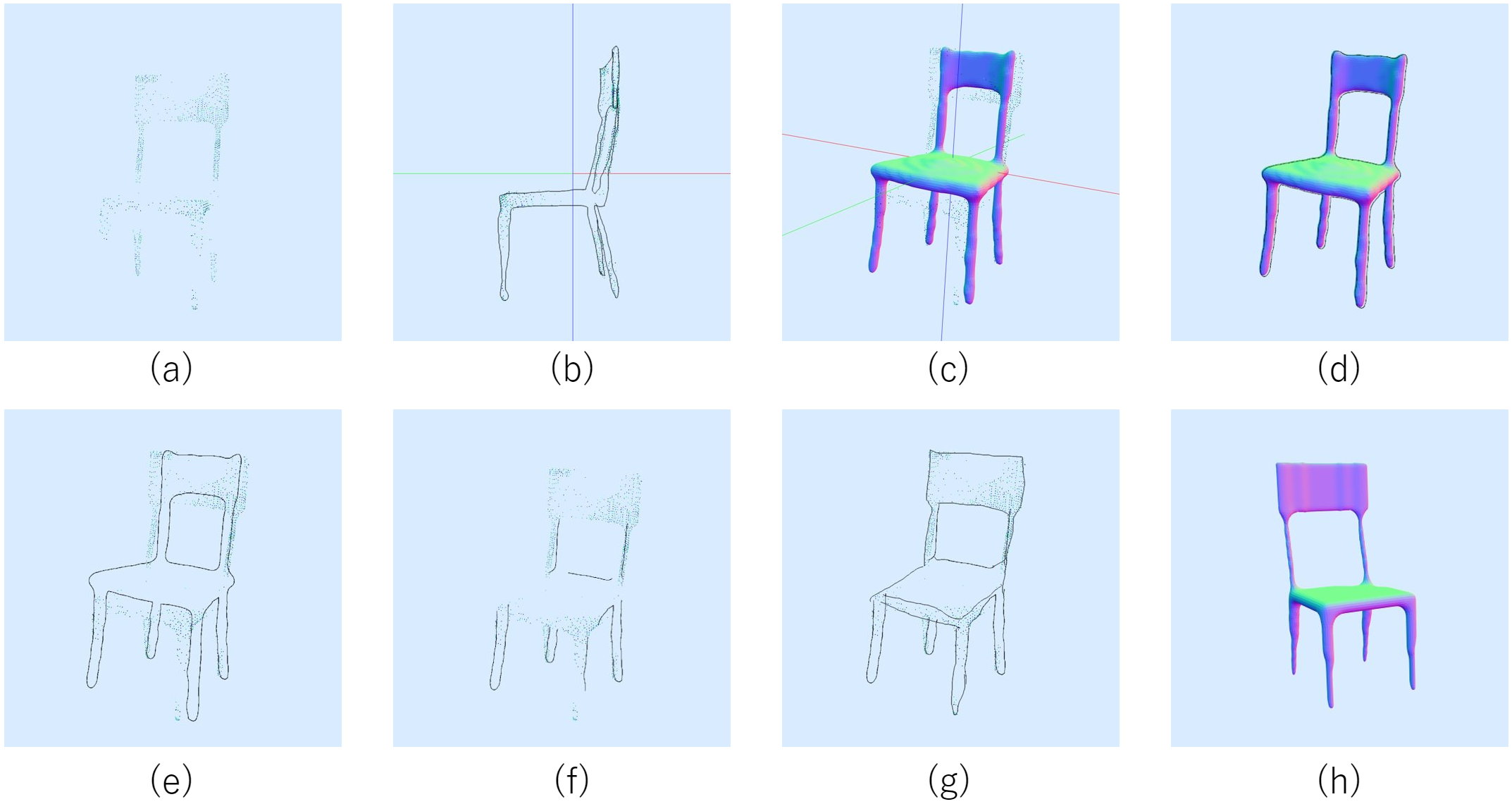}
\end{tabular}
\end{center}
\caption[example] 
{ \label{fig:result-process} 
Images of the user's experiment following our proposed complete process:
(a) sparse point cloud,
(b) sketch drawn,
(c) point cloud and mismatched model after performing ICP alignment,
(d) contours of the current model,
(e)—(g) modification of the model contour to match the point cloud features,
(h) target model.}
\end{figure} 

Figure 5 shows that the user retrieves a model that does not match the target and then uses the ICP algorithm to align this model with the point cloud and extract the contour of the model. The user can modify and edit this contour to match the features of the point cloud and use the modified sketch for the next retrieval step. The method we provide allows the user to find the target model faster and more accurately, based on previous work. It is also more in accordance with the user's habit of modifying and editing the previous sketch.

\begin{table}[ht]
\caption{Evaluation results from our user study.} 
\label{tab:Multimedia-Specifications}
\begin{center}       
\begin{tabular}{|c|c|c|c|c|c|}
\hline
&Item & Min & Max & Average & Standard Deviation  \\
\hline
Objective Evaluation&Times of Sketching & 1 & 5 & 1.67 & 1.31   \\
\cline{2-6}
&Sketch Similarity & 0.42 & 0.76 & 0.61 & 0.10  \\
\cline{2-6}
&ICP error & $4.00 \times 10^{-2}$ & $6.67 \times 10^{-2}$ & $5.37 \times 10^{-2}$ & $7.50 \times 10^{-3}$  \\
\hline
Subjective Evaluation&User Satisfaction & 3 & 4 & 3.83 & 0.37   \\
\cline{2-6}
&Easiness to Use & 2 & 5 & 4.33 & 1.11  \\
\hline
\end{tabular}
\end{center}
\end{table}

\subsection{Evaluation Results}
We collected and analyzed data from participants during the user study. As shown in Table 1, the average number of times they needed to draw was 1.67. The mean similarity of the sketch drawn by the user was 0.61. The mean value of ICP error was $5.37 \times 10^{-2}$. The ICP alignment results were relatively stable, and the point cloud and the retrieved model can overlap to a high degree. It is verified that the proposed approach enables users to retrieve the target model in less time, at less cost, and more accurately.

After the user study, we asked all participants to fill out a questionnaire to verify the usefulness and user experience of the proposed interface. We invited users to rate ``how satisfied they were with obtaining the model" and ``how easy it was to use the interface" with 5-point Likert scales (1 = strongly disagree, 5 = strongly agree). As shown in Table 1, users rated the acquired model with a mean satisfaction score of 3.83 and standard deviation was 0.37. Most of the participants were relatively satisfied with the acquired models. Users rated the ease of use of the interface with a mean score of 4.33, and the standard deviation was 1.11. Most users thought our interface was easy to use.  The users' feedback confirmed that the proposed interface can obtain the target model quickly and efficiently based on the sketch. It was easy to recognize different angles of view without being limited to their recognition abilities. Some users also felt that our user interface was close to the real application scenario and the user's drawing habits.

\section{Conclusion}

We proposed a sketch-based human-in-the-loop interactive system to retrieve and reconstruct 3D models from sparse point clouds. We provided an iterative strategy to help users modify and improve the sketch. Benefiting from the freely observable point cloud data, users can draw sketches of structural contours in arbitrary viewpoints with a free proficiency level and imagination. We conducted a user study and found that the target 3D model can be acquired well by drawing sketch contours from sparse point clouds. We plan to expand the variety of objects in the dataset with complex objects, such as the human body. Furthermore, we would like to explore the deep learning-based generative model for shape generation rather than model retrieval.

\acknowledgments % equivalent to \section*{ACKNOWLEDGMENTS}    
We thank all the participants in our user study. This work was supported by JAIST Research Grants, and JSPS KAKENHI Grant 20K19845, Japan.

% References
\bibliography{report} % bibliography data in report.bib

\begin{thebibliography}{10}

\bibitem{b1}
Maghoumi, M., LaVioia, J.~J., Desingh, K., and Chadwicke~Jenkins, O.,
  ``Gemsketch: Interactive image-guided geometry extraction from point
  clouds,'' in [{\em 2018 IEEE International Conference on Robotics and
  Automation (ICRA)}{\nolinebreak\hspace{0.1em}]},   2184--2191 (2018).

\bibitem{b17}
Berger, M., Tagliasacchi, A., Seversky, L.~M., Alliez, P., Guennebaud, G.,
  Levine, J.~A., Sharf, A., and Silva, C.~T., ``A survey of surface
  reconstruction from point clouds,'' {\em Computer Graphics Forum}~{\bf
  36}(1),  301--329 (2017).

\bibitem{b15}
Nan, L. and Wonka, P., ``Polyfit: Polygonal surface reconstruction from point
  clouds,'' in [{\em Proceedings of the IEEE International Conference on
  Computer Vision (ICCV)}{\nolinebreak\hspace{0.1em}]},  (Oct 2017).

\bibitem{b16}
Brüel-Gabrielsson, R., Ganapathi-Subramanian, V., Skraba, P., and Guibas,
  L.~J., ``Topology-aware surface reconstruction for point clouds,'' {\em
  Computer Graphics Forum}~{\bf 39}(5),  197--207 (2020).

\bibitem{b8}
Eitz, M., Richter, R., Boubekeur, T., Hildebrand, K., and Alexa, M.,
  ``Sketch-based shape retrieval,'' {\em ACM Trans. Graph. (Proc.
  SIGGRAPH)}~{\bf 31}(4),  31:1--31:10 (2012).

\bibitem{peng21}
Peng, Y., Huang, Z., Zhao, C., Xie, H., Fukusato, T., and Miyata, K.,
  ``Sketch-based human motion retrieval via shadow guidance,'' in [{\em 2021
  Nicograph International (NicoInt)}{\nolinebreak\hspace{0.1em}]},   42--45
  (2021).

\bibitem{he2021}
He, Y., Xie, H., Zhang, C., Yang, X., and Miyata, K., ``{Sketch-based normal
  map generation with geometric sampling},'' in [{\em International Workshop on
  Advanced Imaging Technology (IWAIT) 2021}{\nolinebreak\hspace{0.1em}]},
  Nakajima, M., Kim, J.-G., Lie, W.-N., and Kemao, Q., eds.,  {\bf 11766},  261
  -- 266, International Society for Optics and Photonics, SPIE (2021).

\bibitem{dualface2021}
Huang, Z., Peng, Y., Hibino, T., Zhao, C., Xie, H., Fukusato, T., and Miyata,
  K., ``dualface: Two-stage drawing guidance for freehand portrait sketching,''
  {\em Computational Visual Media}~{\bf 8},  63–77 (2022).

\bibitem{xie21}
Xie, H., Peng, Y., Wang, H., and Miyata, K., ``Sketchmehow: Interactive
  projection guided task instruction with user sketches,'' in [{\em HCI
  International 2021 - Late Breaking Papers: Cognition, Inclusion, Learning,
  and Culture}{\nolinebreak\hspace{0.1em}]},   513--527, Springer International
  Publishing, Cham (2021).

\bibitem{b3}
Chen, T., Zhu, Z., Shamir, A., Hu, S.-M., and Cohen-Or, D., ``3-sweep:
  Extracting editable objects from a single photo,'' {\em ACM Trans.
  Graph.}~{\bf 32} (nov 2013).

\bibitem{b4}
Maghoumi, M., LaVioia, J.~J., Desingh, K., and Chadwicke~Jenkins, O.,
  ``Gemsketch: Interactive image-guided geometry extraction from point
  clouds,'' in [{\em 2018 IEEE International Conference on Robotics and
  Automation (ICRA)}{\nolinebreak\hspace{0.1em}]},   2184--2191 (2018).

\bibitem{dataset12}
Li, B. and Godil, A., ``Shrec12 track: Sketch-based 3d shape retrieval,''
  Eurographics Workshop on 3D Object Retrieval (2012), Cagliari (2012-05-13
  2012).

\bibitem{b20}
Zhang, D., ``Opensse: Open sketch search engine.''
  \url{https://github.com/zddhub/opensse}.

\bibitem{b21}
Zhou, Q.-Y., Park, J., and Koltun, V., ``{Open3D}: {A} modern library for {3D}
  data processing,'' {\em arXiv:1801.09847}  (2018).

\end{thebibliography}
\bibliographystyle{spiebib} % makes bibtex use spiebib.bst

\end{document}